# New stable two dimensional silicon carbide nanosheets

**Qun Wei[1], Ying Yang[2, 3], Guang Yang[2, 4], Xihong Peng[2*],**

[1]School of Physics and Optoelectronic Engineering, Xidian University, Xi'an, Shaanxi 710071, P. R. China

[2]College of Integrative Sciences and Arts, Arizona State University, Mesa, Arizona 85212, USA

[3]School of Automation and Information Engineering, Xi'an University of Technology, Xi'an, Shaanxi, 710048, P. R. China

[4]School of Science, Hebei University of Science and Technology, Shijiazhuang, Hebei 050018, P. R. China

## ABSTRACT

We predict the existence of new two dimensional silicon carbide nanostructure employing *ab initio* density-functional theory calculations. These structures are composed of tetragonal and hexagonal rings with C-C and Si-C bonds arranged in a buckling plane. They are proven to be thermodynamically and mechanically stable with relatively low formation energy, implying potential fabrication in lab. They exhibit strong ductility and anisotropicity from their strain-stress relation and directional dependence of mechanical moduli. The materials maintain phonon stability upon the application of mechanical strain up to 27% with fantastic ductile property. The $SiC_2$ structure possesses a tiny direct band gap of 0.02 eV predicted using HSE06 functional and the band gap can be opened up through multiple approaches such as hydrogenation and strain application. The gap values can be strategically tuned in the range of 0.02 ~ 1.72 eV and the direct/indirect gap nature can be further manipulated. In contrast, a closely related structure of SiC shows an indirect HSE band gap of 1.80 eV and strain engineering its value between 0.0 ~ 1.95 eV. The unique properties in these newly proposed structures might have potential applications in future nanomechanics and electronics.

---

[*] To whom correspondence should be addressed.  E-mail: xihong.peng@asu.edu.

# 1. Introduction

Successful fabrication of two dimensional (2D) materials such as group IV graphene [1–3] and silicene [4–6] prompt incredible interests in 2D materials research. Graphene was explored enormously in practical applications [1–3]. Silicene embracing graphene-like crystal lattice with a low-buckled geometry was also been extensively studied [4–6] due to its compatibility and desirability in device applications [6]. However, both graphene and silicene are semimetals with π and π* states touching each other at the Fermi level [7]. The zero band gap in graphene and silicene limits their widespread applications in optoelectronic devices such as light-emitting diodes, field effect transistors, and solar cells. Numerous strategies have been investigated to open the band gap in graphene and silicene, including substrate effects [4–6], nanoribbons with quantum confinement [8, 9], alloy [10], applying external electric fields [11], hydrogenation [12], chemical functionalization [13], doping [14], introducing defects [15] and applying mechanical strain [16]. Another way to engineer desirable band gap is to explore other 2D materials. In addition to graphene (silicene), various other 2D carbon (silicon) allotropes, such as penta-graphene [17], T-graphene [18], S-graphene [19], $\alpha$-, $\beta$-, $\delta$-graphynes [20, 21], tetrahex-carbon [22] were further studied and the properties of each unique structure were reported in literature.

For silicon-carbon binary compounds, silicon carbide is known to have more than 250 crystalline polytypisms including cubic SiC, 4H- and 6H-SiC. Silicon carbide compounds traditionally have four-coordinated tetragonal building blocks with remarkable electronic properties and superior thermal, chemical, and mechanical stability. Electronic devices based upon these properties have been utilized at temperatures in excess of 300 °C and in harsh environments. Recently, numerous 2D silicon carbide compounds were also fabricated and proposed, for instance, experimentally synthesized graphitic SiC [4–6], planar $SiC_2$ silagraphene with tetra-coordinated Si [23], planar graphitic $SiC_2$ [24], carbon-rich $SiC_3$ [25], g-$SiC_2$ [24], pt-$SiC_2$ [23], $SiC_6$-SW [26], $SiC_2$-b [26], $SiC_2$-p [26], quasi-planar tetragonal SiC and $SiC_2$ [27], penta-$SiC_2$ [28], a series of silagraphyne [27], silicon-rich $Si_3C$ [26], and recently reported tetrahex SiC [29]. These 2D nanostructures exhibit either metallic (semimetal) or semiconducting with a finite band gap.

In this work, we propose a new type of 2D silicon carbide nanosheet with a buckled geometry. Similar to the structure of tetrahex-carbon [22], the new 2D silicon carbide network contains tetragonal (T) and hexagonal (H) rings, named as TH-$SiC_2$ and TH-SiC. In TH-$SiC_2$, the four-coordinated ($sp^3$) Si atoms are sandwiched between two sublayers of three-coordinated ($sp^2$) carbon atoms. In TH-SC, both Si and C atoms can be $sp^3$ or $sp^2$ hybridized. Through the calculations of first-principles density-functional theory (DFT) [30], these new nanostructures are proven to be thermodynamically and mechanically stable. They have relatively low formation energy and indicates potential fabrication in lab. In the frame work of DFT, we find that that TH-SiC$_2$ possesses a tiny band gap 0.02 eV predicted by the hybrid Heyd-Scuseria-Ernzerhof (HSE)06 method [31, 32], while TH-SiC has a finite indirect gap of 1.80 eV. These materials are ductile to sustain a high strain up to 27% while still maintaining phonon stability. Their band gap can be effectively manipulated through various strategies including hydrogenation and mechanical strain application.

## 2. Computational details

The first-principles DFT [30] calculations are carried out using VASP package [33, 34] with projector-augmented wave (PAW) potentials [35, 36]. Perdew-Burke-Ernzerhof (PBE) exchange-correlation functional [37] is selected for geometry relaxation. The hybrid (HSE)06 method [31, 32] is used to calculate electronic band structure and band gap due to its improved accuracy on predicting semiconductor band gaps.

Wave functions of valence electrons are described using a plane wave basis set with kinetic energy cutoff 900 eV. The reciprocal space is meshed 15 × 13 × 1 by using Monkhorst-Pack method [38]. A conjugate-gradient algorithm [39] is used to relax the ions into their instantaneous ground state. The energy convergence criterion for electronic self-consistent procedure is set to be $10^{-8}$ eV and the force is converged to be less than $10^{-4}$ eV/ Å for geometry optimization of the simulation cell. The kinetic energy cutoff 500 eV for the plane wave basis set is utilized for the HSE band structure calculations. A total of 11 *k*-points are collected along each high symmetry line in the reciprocal space for the band structure calculations. The **c**-vector of the unit cell is set to be 20 Å to ensure sufficient vacuum space (> 16 Å) included in the calculations to minimize the interaction between the system and its replicas resulted from periodic boundary conditions. Phonon spectrum is conducted by density functional perturbation theory method using a supercell approach in the PHONOPY code [40] with the forces computed from VASP [33, 34].

## 3. Results and discussion

### 3.1. Structure and thermodynamic stability

Fig. 1(a) and 1(b) presents the crystal structures of the newly predicted 2D TH-SiC$_2$ and its related TH-SiC. They are buckled three-sublayer crystal structure with four-coordinated ($sp^3$) atoms being sandwiched between two sublayers of three-coordinated ($sp^2$) atoms. The rectangular conventional cell contains 12 atoms, 4 Si and 8 C atoms in TH-SiC$_2$ and 6 Si and 6 C in SiC. The buckling thickness $d$ describes the distance between the top and bottom sublayers of the $sp^2$-atoms. The length $r_1$ is the Si-C bond in the tetragonal rings and $r_2$ is the C-C bond (Si-C bond) in the hexagonal rings for the TH-SiC$_2$ (SiC) structure.

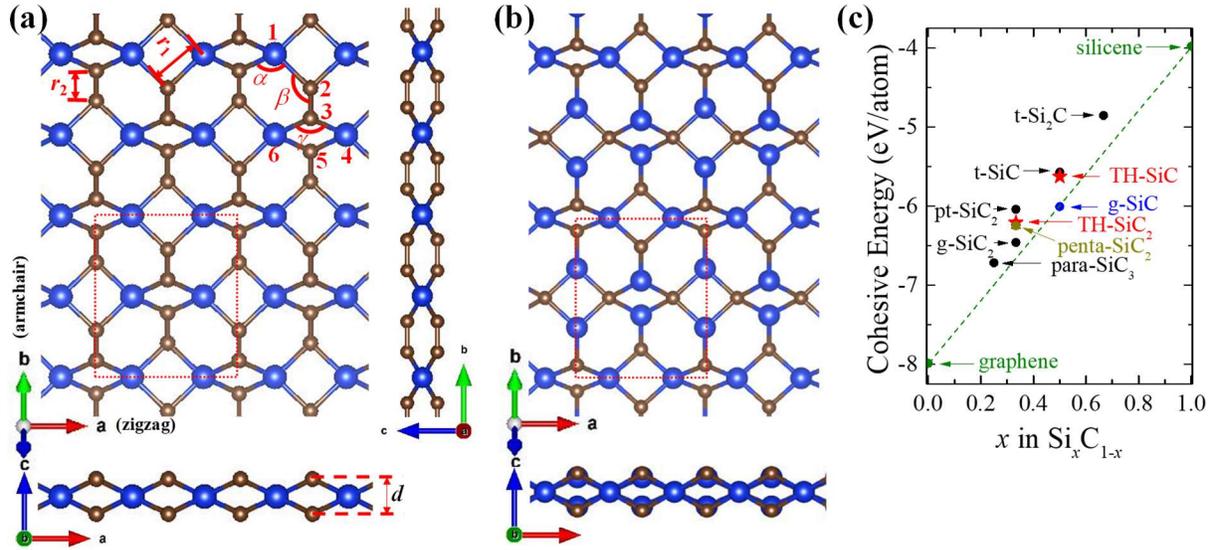

Fig. 1. Snapshots of the 2D (a) TH-SiC$_2$ and (b) SiC structure viewed from various orientations. Si and C atoms are in blue and brown, respectively. The buckling thickness is labeled as $d$. The bond lengths, bond angles are denoted. The dashed rectangle represents a conventional cell containing 12 atoms. (c) Cohesive energy of some 2D Si$_x$C$_{1-x}$ compounds. The green dashed line connecting the energies of graphene and silicene is a guide for eye. For structural details, one can refer to pertinent papers (para SiC$_3$ [25], penta-SiC$_2$ [28], g-SiC [41], g-SiC$_2$ [24], pt-SiC$_2$ [23], t-SiC [27], t-Si$_2$C [27]).

The relaxed lattice constants for the 2D TH-SiC$_2$ are $a = 5.52$ Å and $b = 7.20$ Å. The buckling thickness $d$ is 1.34 Å. The bond lengths $r_1$ and $r_2$ are predicted to be 1.90 Å and 1.35 Å, respectively. The calculated bond angles denoted in Fig. 1(a) are $\alpha = 107.5°$, $\beta = 126.3°$, and $\gamma = 92.9°$. It is clear from the tilt-axis view in Fig. 1(a), two neighbored hexagonal rings along the **a**-axis ($x$) are not coplanar. The dihedral angle between these two hexagonal rings is denoted as $\phi_{1234}$ by the neighboring atoms 1-2-3-4 and it is predicted to be $\phi_{1234} = 128.1°$. Similarly, the neighbored hexagonal and tetragonal rings are neither coplanar and its dihedral angle $\phi_{2345} = 140.6°$. Similarly, for the SiC structure in Fig. 1(b), the corresponding geometry parameters are $a = 5.53$ Å, $b = 7.63$ Å, $d = 1.60$ Å, $r_1 = 1.88$ Å, $r_2 = 1.72$ Å, $\alpha = 116.1°$, $\beta = 118.9°$, $\gamma = 94.6°$, $\phi_{1234} = 126.8°$, $\phi_{2345} =$ 149.5°. These parameters on the TH-SiC structure are in great agreement with a recently reported work [29].

Table 1 gives a summary of some structural parameters and basic properties of the TH-SiC$_2$ and SiC, along with penta-SiC$_2$ [28], tetrahex-carbon [22, 42], penta-graphene [17], graphene and low-buckled silicene [7, 27]. Comparing the buckling thickness and two typical bond lengths in TH-SiC$_2$ with those in penta-SiC$_2$, their values are close. While for TH-SiC, the bond $r_2$ is significantly lengthened since it describes Si-C bonds in SiC instead of C-C as in SiC$_2$. The buckling thickness $d$ is also increased indicating the structure of SiC is more buckled compared to TH-SiC$_2$ and penta-SiC$_2$.

*Table 1. Summary of structural parameters and basic properties of TH-SiC$_2$, SiC, penta-SiC$_2$, tetrahex-C, penta-graphene, graphene, and silicene. Lattice constants a, b, buckling thickness d, and bond lengths $r_1$, $r_2$ are in unit of Å, cohesive $E_{coh}$ and formation energies $E_{form}$ are in eV/atom, HSE predicted band gap $E_g$ in eV. The formation energy is calculated using graphene and silicene as references (thus \*zero for graphene and silicene). Reference: TH-SiC [29], penta-SiC$_2$ [28], tetrahex-C [22, 42], penta-graphene [17], silicene [7, 27].*

| Structures | a | b | d | $r_1$ | $r_2$ | $E_{coh}$ | $E_{form}$ | $E_g$ | Gap nature |
|---|---|---|---|---|---|---|---|---|---|
| TH-SiC2 | 5.52 | 7.20 | 1.34 | 1.90 | 1.35 | -6.21 | 0.44 | 0.02 | direct |
| TH-SiC | 5.53 | 7.63 | 1.60 | 1.88 | 1.72 | -5.63 | 0.61 | 1.80 | indirect |
| Penta-SiC2 | 4.41 | 4.41 | 1.33 | 1.91 | 1.36 | -6.25 | 0.40 | 2.85 | indirect |
| Tetrahex-C | 4.53 | 6.10 | 1.16 | 1.53 | 1.34 | -7.12 | 0.87 | 2.64 | direct |
| Penta-graphene | 3.64 | 3.64 | 1.20 | 1.55 | 1.34 | -7.09 | 0.91 | 3.25 | indirect |
| Graphene | 2.46 | 2.46 | 0.00 | 1.42 | 1.42 | -7.99 | 0* | 0.00 | |
| Silicene | 3.87 | 3.87 | 0.44 | 2.28 | 2.28 | -3.98 | 0* | 0.00 | |

Cohesive energy $E_{coh}$ for a general silicon carbide compound Si$_x$C$_y$ is computed using the following equation,

$$E_{coh} = \frac{E_{tot}(\text{Si}_x\text{C}_y) - x\,E_{Si} - y\,E_C}{x+y}, \qquad (1)$$

where $E_{tot}$, $E_{Si}$, $E_C$ are the total energy of the material, and the energies of an isolated silicon and carbon atoms, respectively. The formation energy $E_{form}$ per atom can then be obtained as follows [43, 44],

$$E_{form} = E_{coh}(Si_xC_y) - \frac{x}{x+y}\mu_{Si} - \frac{y}{x+y}\mu_C, \text{ or } E_{form} = \frac{E_{tot}(Si_xC_y) - x\,E(Si) - y\,E(C)}{x+y}, \quad (2)$$

where $\mu_{Si}$, $\mu_C$, are cohesive energy per atom in silicene and graphene, respectively, $E(Si)$, $E(C)$ are the total energies per atom in silicene and graphene, respectively. Note that $\mu_{Si} = E(Si) - E_{Si}$ and $\mu_C = E(C) - E_C$.

The calculated cohesive and formation energies are provided in Table 1. Fig. 1(c) plots their cohesive energy along with some other 2D $Si_xC_{1-x}$ compounds. The green dashed line connecting the energies of graphene and silicene is used to gauge the formation energy. The formation energy is positive (negative) above (below) the line. For structural details, one can refer to pertinent papers: para $SiC_3$[25], penta-$SiC_2$ [28], g-SiC [41], g-$SiC_2$ [24], pt-$SiC_2$ [23], t-SiC [27], t-$Si_2C$ [27].

It is found that in Fig. 1(c) TH-$SiC_2$ has lower cohesive energy by 0.17 eV/atom compared to the planar pt-$SiC_2$ [23]. It has slightly higher cohesive energy by 0.04 eV/atom, compared to penta-$SiC_2$. This is opposite to their carbon counterparts since tetrahex-C is slightly energetically more favorable than penta-graphene (i.e. -7.12 versus -7.09 eV/atom). The formation energy of TH-$SiC_2$ is found to be 0.44 eV/atom, which is 0.04 eV/atom higher that penta-$SiC_2$. On the other hand, TH-SiC has a lower cohesive energy by 0.06 eV/atom compared to t-SiC [27], while a higher energy by 0.38 eV/atom to g-SiC [41]. The positive formation energy in TH-$SiC_2$ and SiC indicates that their structures are metastable, similar to many other $Si_xC_{1-x}$ compounds as presented in in Fig. 1(c).

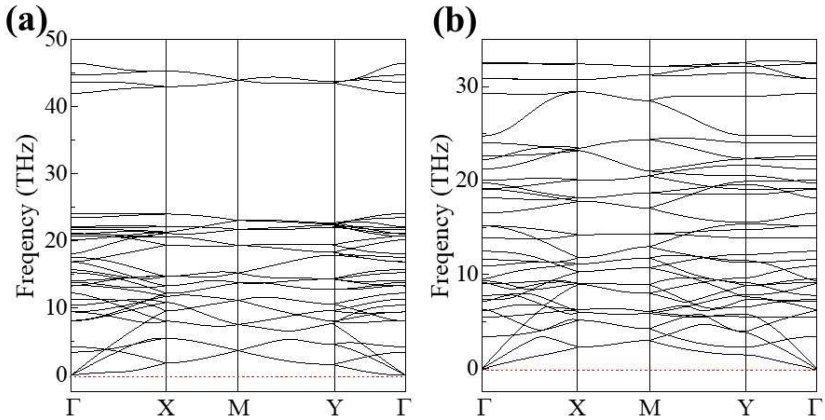

Fig. 2. Phonon spectra of the 2D (a) TH-$SiC_2$ and (b) SiC nanosheets.

The TH-$SiC_2$ and SiC structures are proven to be thermodynamically stable according to the phonon spectrum calculations as shown in Fig. 2. No imaginary frequencies are found in the phonon spectra for the structures, indicating that they are stable. The stability of the structures under mechanical strain are also explored. Starting with the fully relaxed 2D crystal structures of TH-SiC$_2$ and SiC, biaxial and uniaxial tensile strain up to 40% at an increment of 1% is applied in either the *x* (**a** or zigzag) or *y* (**b** or armchair) direction. The tensile strain is defined as,

$$\varepsilon = \frac{a-a_0}{a_0} \quad (3)$$

where $a$ and $a_0$ are the lattice constants of the strained and relaxed structures, respectively. In the case of uniaxial strain applied in one direction, the lattice constant in the transvers direction is fully relaxed to ensure minimal stress in the transverse direction.

The phonon spectra of the materials under various values of strain are shown in Fig. 3. No negative frequency is observed in the top row of the spectra. However, negative frequencies appear in the bottom row, indicating instability of the structures. TH-SiC$_2$ structure remains stable up to 27%, 9% and 16% for the uniaxial strain in the zigzag, armchair directions, and biaxial strain, respectively. While SiC is stable up to 15% and 16% for the uniaxial strain in the armchair direction and biaxial strain, respectively. Higher strain than those values bring in negative frequencies in the spectra and suggest that the structures are no longer stable. Interestingly, SiC is unstable for any value of strain applied in the *x*-direction.

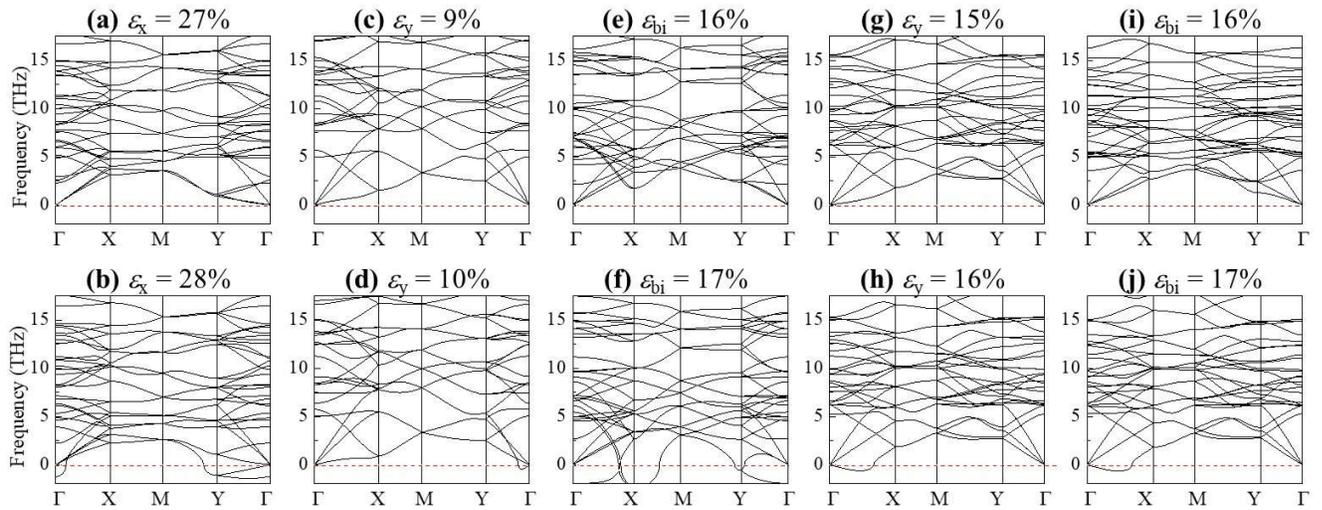

Fig. 3. Phonon spectra in the strained 2D (a)-(f) TH-SiC$_2$ and (g)-(j) SiC nanostructures. The appearance of negative frequencies at the bottom row indicates structural instability.

### 3.2. Mechanical properties and mechanical stability

The mechanical properties and stability of the new structure are also studied thoroughly. Study of strain-stress relation in a material can determine its ideal strength (the highest strength of a crystal at 0 K) [45, 46] and critical strain (at which ideal strength reaches) [47]. To obtain the strain-stress relation in the 2D TH-SiC$_2$ and SiC structures, uniaxial tensile strain along both $x$ (zigzag) and $y$ (armchair) directions and biaxial strain are applied to the materials with an increment of 1% up to 40%. The results are given in Fig. 4. The strain-stress relation is calculated using the method described in the references [48, 49], which was designed for three dimensional (3D) material. For a 2D system, the stress calculated from the DFT calculations has to be adjusted since DFT reported stress is largely underestimated due to averaging force over vacuum space. To avoid this effect, the stress in this work adopts the force per unit length in the unit of N/m. It is found that TH-SiC$_2$ and SiC are more ductile in the $x$ (zigzag) direction, similar to the case of tetrahex-C [42].

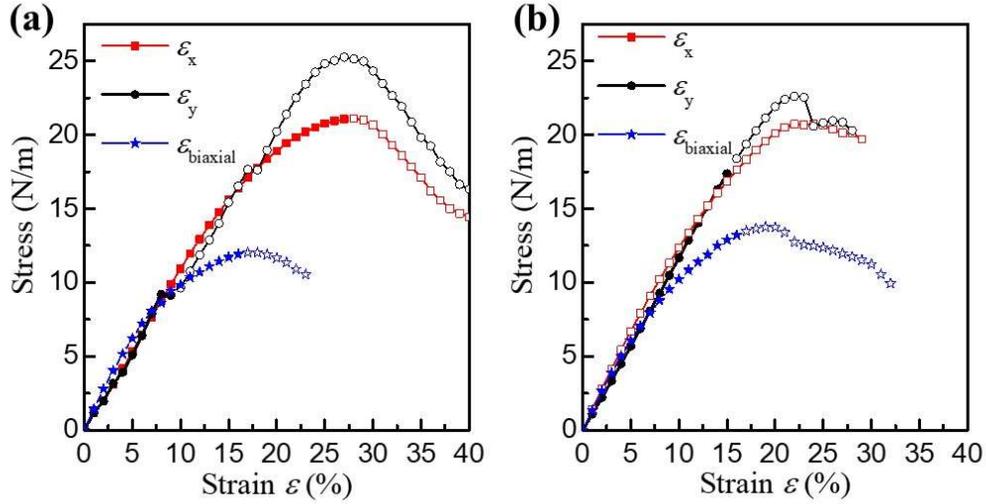

*Fig. 4. The strain-stress relation in the (a) TH-SiC$_2$ and (b) TH-SiC for uniaxial strain applied in the zigzag (x), armchair (y), and biaxial directions, respectively. Phonon instability occurs for TH-SiC$_2$ when strain is beyond 27%, 9%, and 16% in the x, y, and biaxial directions with the corresponding strength 21.1, 9.1, and 11.9 N/m, respectively. For TH-SiC, phonon instability occurs when strain is beyond 15% in the y axis and 16% biaxial strain with corresponding strength 17.4 and 13.2 N/m, respectively. TH-SiC is found not stable for uniaxial strain applied in the zigzag direction. Solid and hollow symbols represent phonon stable and instable, respectively.*

As mentioned previously, phonon instability occurs in TH-SiC$_2$ when strain is beyond 27%, 9%, and 16% in the $x$, $y$, and biaxial directions, respectively. Solid symbols in Fig. 4 represent stable structures and hollow symbols for instable structure. Consequently, the ideal strength of TH-SiC$_2$ is found to be 21.1, 9.1 and 11.9 N/m in the $x$, $y$ and biaxial direction, respectively. This high strength of 21.1 N/m is close to that in penta-graphene (23.5 N/m strength with 18% uniaxial strain in both zigzag and armchair directions) [50]. However, it is lower than that of graphene [46] and tetrahex-C [42]. On the other hand in TH-SiC, phonon instability occurs when strain is beyond 15% in the *y* axis and 16% biaxial strain with corresponding strength 17.4 and 13.2 N/m, respectively. This structure is found not stable at all for any uniaxial strain applied in the zigzag direction. Our reported strength values in TH-SiC are smaller than those in the reference [29], since the referenced work does not consider phonon stability for the strained structure. Its reported ultimate tensile strength values 25, 23 and 21 N/m for *x*, *y*, and biaxial strains, respectively, occur at the strains which we have proved that the phonon spectrum of the structure is not stable as shown in Fig. 3.

The energy surface of the 2D TH-SiC$_2$ and SiC structures are scanned in the small strain range -0.6% < $\varepsilon_{xx}$ < +0.6%, -0.6% < $\varepsilon_{yy}$ < +0.6%, and -0.6% < $\varepsilon_{xy}$ < +0.6% to calculate their elastic stiffness constants and various moduli. The strain energy is defined as the energy difference between the strained and relaxed structures,

$$E_s = E(\varepsilon) - E_0 \tag{4}$$

where $E(\varepsilon)$ and $E_0$ are the total energy of strained and relaxed structures, respectively. The calculated strain energy is then fitted parabolically using the following equation, to determine the coefficients $a_i$, and the elastic stiffness constants are readily calculated as,

$$E_s = a_1 \varepsilon_{xx}^2 + a_2 \varepsilon_{yy}^2 + a_3 \varepsilon_{xx} \varepsilon_{yy} + a_4 \varepsilon_{xy}^2 \tag{5}$$

$$C_{ij} = \frac{1}{A_0} \left( \frac{\partial E_s^2}{\partial \varepsilon_i \varepsilon_j} \right), \tag{6}$$

where *i, j* = *xx*, *yy*, or *xy*, $A_0$ is the area of the simulation cell in the *xy* plane. The Young's and shear moduli for a 2D system can be derived as a function of $a_i$ [42, 47],

$$E_x = \frac{4a_1 a_2 - a_3^2}{2 a_2 A_0}, E_y = \frac{4a_1 a_2 - a_3^2}{2 a_1 A_0}, G_{xy} = \frac{2a_4}{A_0}. \tag{7}$$

When uniaxial strain is applied in one direction, the lattice constant in the transvers direction need to be fully relaxed to ensure minimal stress in the transverse direction. And the response strain in the transverse direction can be also calculated according to Eq. (3). Poisson's ratio can then being readily calculated according to its definition,

$$v = -\frac{\varepsilon_{transverse}}{\varepsilon_{axial}}, v_{xy} = -\frac{\varepsilon_y}{\varepsilon_x}, v_{yx} = -\frac{\varepsilon_x}{\varepsilon_y} \tag{8}$$

where $\varepsilon_{axial}$ and $\varepsilon_{transverse}$ are the applied axial strain and its response strain in the transverse direction, respectively.

The calculated mechanical parameters in TH-SiC$_2$ and TH-SiC are listed in Table 2. The Poisson's ratios in TH-SiC is significantly smaller than those in TH-SiC$_2$ due to its smaller value of C$_{12}$. The elastic constants can be used to check the mechanical stability of the structures. The positive values in the elastic matrix in TH-SiC2 and TH-SiC (see Table 2) shows the mechanical stability of these new structures. Our predicted parameters for TH-SiC are in great consistence with a recently reported work [29].

Table 2. A summary of mechanical parameters in TH-SiC$_2$ and TH-SiC. The elastic stiffness constants $C_{ij}$, Young's moduli E along the x and y directions, and the shear modulus $G_{xy}$ are in unit of N/m.

| Structures | $C_{11}$ | $C_{22}$ | $C_{33}$ | $C_{12}$ | $E_x$ | $E_y$ | $G_{xy}$ | $v_{xy}$ | $v_{yx}$ |
|---|---|---|---|---|---|---|---|---|---|
| TH-SiC$_2$ | 119.7 | 123.3 | 57.3 | 26.5 | 114.0 | 117.4 | 57.3 | 0.22 | 0.22 |
| TH-SiC | 143.7 | 107.7 | 41.8 | 2.8 | 143.6 | 107.6 | 41.8 | 0.025 | 0.019 |

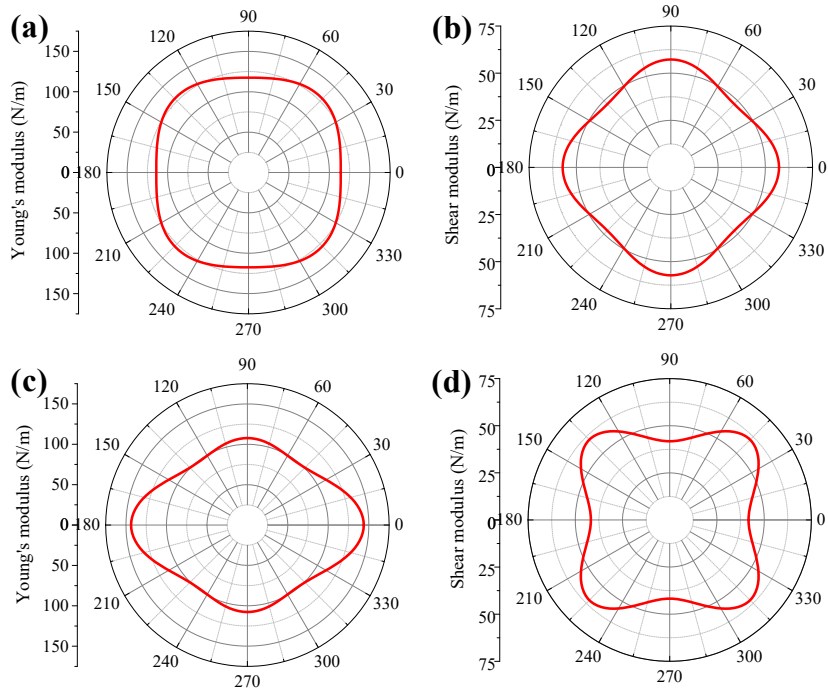

Fig. 5. The directional dependence of (a)(c) Young's modulus and (b)(d) shear modulus in the TH-SiC$_2$ (top) and SiC (bottom).

Young's and shear moduli along an arbitrary direction can be calculated using the methods described in [47] with the angle of an arbitrary direction from the +x axis. The direction dependence of the Young's and shear moduli in TH-SiC$_2$ and TH-SiC are presented in Fig. 5. For TH-SiC$_2$, the maximal Young's modulus is along the [11] direction (i.e. 45° from the *x* axis) with a value of 129 N/m, while a minimum of 114 N/m is along the *x* axis as shown in Fig. 5(a). However, it is opposite for TH-SiC. The Young's modulus has the minimum value of 99 N/m along the [11] direction while the largest value 144 N/m in the *x* direction as in Fig. 5(c). In the case of shear strain displayed in Fig. 5(b)(d), TH-SiC$_2$ has the maximal shear modulus of 57 N/m in the *x/y* directions, and a minimum value of 47 N/m along the [11] direction. However, for TH-SiC, the minimum 42 N/m occurs in the *x/y* directions while the maximum 60 N/m along the [11] direction. Fig. 5 demonstrates a strong anisotropicity in the 2D TH-SiC$_2$ and SiC structures.

### 3.3. Electronic properties

Electronic band structures of TH-SiC$_2$ and SiC are presented in Fig. 6(a) and 6(d) using the hybrid HSE06 functional. It shows that SiC$_2$ possesses a direct band gap with a tiny value of 0.02 eV while SiC has an indirect band gap with a value of 1.80 eV. The valence band maximum (VBM) and conduction band minimum (CBM) in TH-SiC$_2$ are located at Γ. However, for the SiC structure, VBM is located at X while CBM is at Y.

The electron density contour plots of VBM/CBM in both structures are also presented in Fig. 6. Through an analysis of the *spd*-orbital site projection of the VBM and CBM, it is found that the VBM in SiC$_2$ is dominantly contributed by the $p_x$ orbitals on all 12 atoms (4 Si and 8 C) in the conventional cells as shown in Fig. 6(b). However, the CBM is dominated by the $p_z$ orbitals mainly located on the $sp^2$-hybridized C atoms in Fig. 6(c). On the other hand for the SiC structure, the electron charge of the VBM is dominantly contributed by the $p_z$ orbitals on the $sp^2$ C atoms as shown in Fig. 6(e), while CBM is located on both C and Si atoms in Fig. 6(f).

The tiny band gap 0.02 eV in TH-SiC$_2$ can be open up through different strategies. Here, we present two different means to manipulate its band gap. The first method is through hydrogenation by adsorbing hydrogen atoms on either or both sides of the nanosheet. The results are presented in Fig. 7. It can been seen that hydrogen adsorbed on either or both sides of the nanosheet can open up the band gap significantly in Th-SiC$_2$. In the case of single-side hydrogenation, the band gap is predicted to be 0.89 eV by the HSE functional with both VBM and CBM located at Γ as shown in Fig. 7(a). For the both side H adsorption, the band gap becomes indirect with a gap value of 1.72 eV. The VBM is still located at Γ. However, the CBM shifts from Γ to Y in Fig. 7(e). Fig. 7 also provides the electron density contour plots of the VBM and CBM. It is found that in both cases, VBM shows similar charge distribution in Fig. 7(c) and (g) as in Fig. 6(b), where the state is mainly contributed by the $p_x$ orbitals located on all Si/C atoms in the conventional cell. However, it is distinct in CBM. For half-H passivation, the CBM in Fig. 7(d) is dominated by the $p_z$ orbitals located on the $sp^2$-C atoms. However, for the full-H passivation, the CBM in Fig. 7(h) is primarily contributed by the *s-p* hybridization of the C-H bonds.

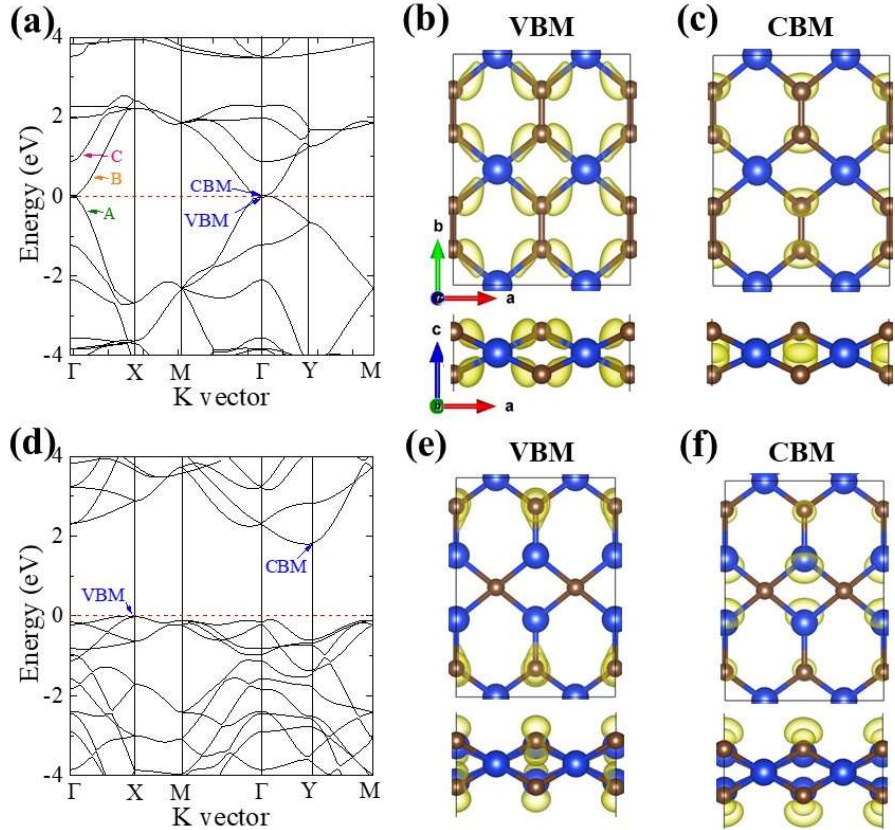

*Fig. 6. (a) The HSE predicted electronic band structure of the relaxed TH-SiC$_2$ and the electron density contour plots of (b) VBM and (c) CBM, respectively. (d)-(f) The band structure and VBM/CBM plots for SiC. The energy of VBM is set at zero. Si and C atoms are in blue and brown, respectively. The isosurface value for the electron density is set at 0.01 e/Bohr$^3$.*

Comparing the band structures in Fig. 6(a) for the pristine TH-SiC$_2$, and Fig 7(a) and 7(e), it can reveal the mechanism for band gap opening through hydrogenation. As shown in Fig. 6(a), the lowest and second lowest conduction bands are labeled as bands B and C, respectively. The electron charge of the band B (i.e. CBM) is primarily contributed by the $p_z$ orbitals located on the $sp^2$-hybridized C atoms as shown in Fig. 6(c). Hydrogen adsorption on the structure converts these $sp^2$ C atoms into $sp^3$ hybridization. This removes the band B and results in the band structure as shown in Fig. 7(a). The second lowest band C is also dominated by the $p_z$ orbitals located on the $sp^2$-C atoms as displayed in Fig. 7(d). Full hydrogen absorption on both sides of the nanosheet completely passivates all $sp^2$-C atoms and converts them into $sp^3$-C, therefore band C also disappears in Fig. 7(e). On the other hand for the valence band, the band A (i.e. VBM) is mainly contributed by the $p_x$ orbitals located at all Si/C atoms including the middle sublayer of $sp^3$-Si atoms. Consequently, this band is not removed through hydrogenation.

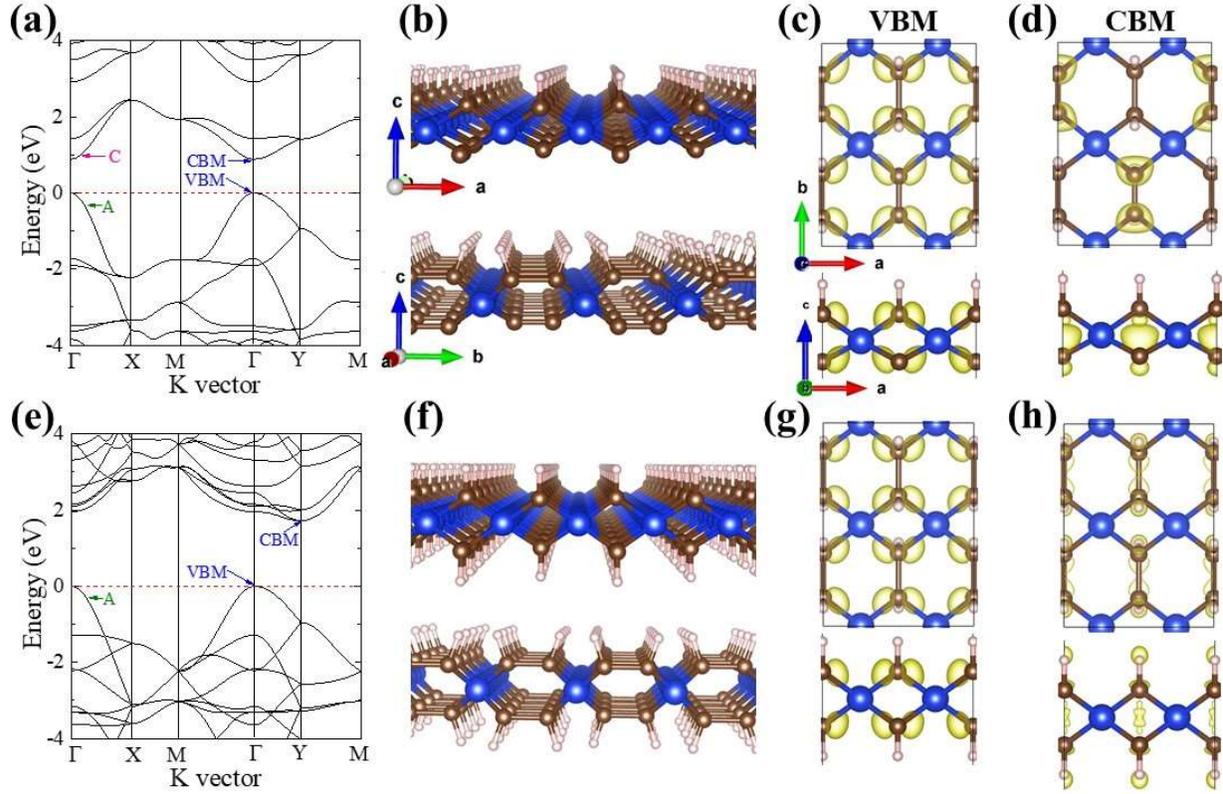

*Fig. 7. Band gap opening through hydrogenation of TH-SiC$_2$. (a)-(d) H adsorbed on one side of the nanosheet, (e)-(h) H adsorbed on both sides of the nanosheet. (a)(e) HSE predicted band structures, (b)(f) schematics of hydrogenation, electron density contour plots of (c)(g) VBM and (d)(h) CBM. Si, C, H atoms are in blue, brown, and grey, respectively.*

The second method to open up the band gap in TH-SiC$_2$ is applying mechanical strain. Figs. 8 and 9 present the strain effects on the electronic properties in SiC$_2$ and SiC, respectively. Fig. 8(a) shows the band structure variation with strain in SiC$_2$. The uniaxial strain is applied in the zigzag ($x$) direction. It is found that strain opens the band gap and maintains the nature of direct gap in the strain range up to 27% till phonon instability occurs. Biaxial and uniaxial strain in the $y$ axis are not able open the band gap. Fig. 8(b) presents the gap variation with the uniaxial strain. The energy variation of the VBM and CBM is given in Fig. 8(c), in which the CBM energy increases while VBM reduces within the strain range till 12%. These different energy variation trends are related to their specific orbitals and bonding/antibonding nature in the $x$ direction. From the charge distribution of the VBM and CBM shown in Fig. 8(d) and 8(e), one can conclude that VBM exhibits antibonding characteristics along the *x* direction while CBM demonstrates bonding nature. With axial strain applied in the *x* direction, the energy variation with strain of the state obeys a pattern schematically illustrated in Fig. 8(f) [51, 52]. This schematics is obtained from the exchange energy model suggested by Heitler-London [53], in which the energies of the bonding and antibonding states are distinct in terms of the exchange-correlation energy of electrons. The exchange-correlation energy is resulted from either non-classical electron-electron (positive) or electron-ion interaction (negative) with the latter dominant for orbitals with a non-localized electron density. Consequently, tensile strain increases bond lengths and the energy of bonding states increases while that of anti-bonding states reduces [51, 52]. The energy variation of VBM and CBM in Fig. 8(c) is beautifully explained by this model.

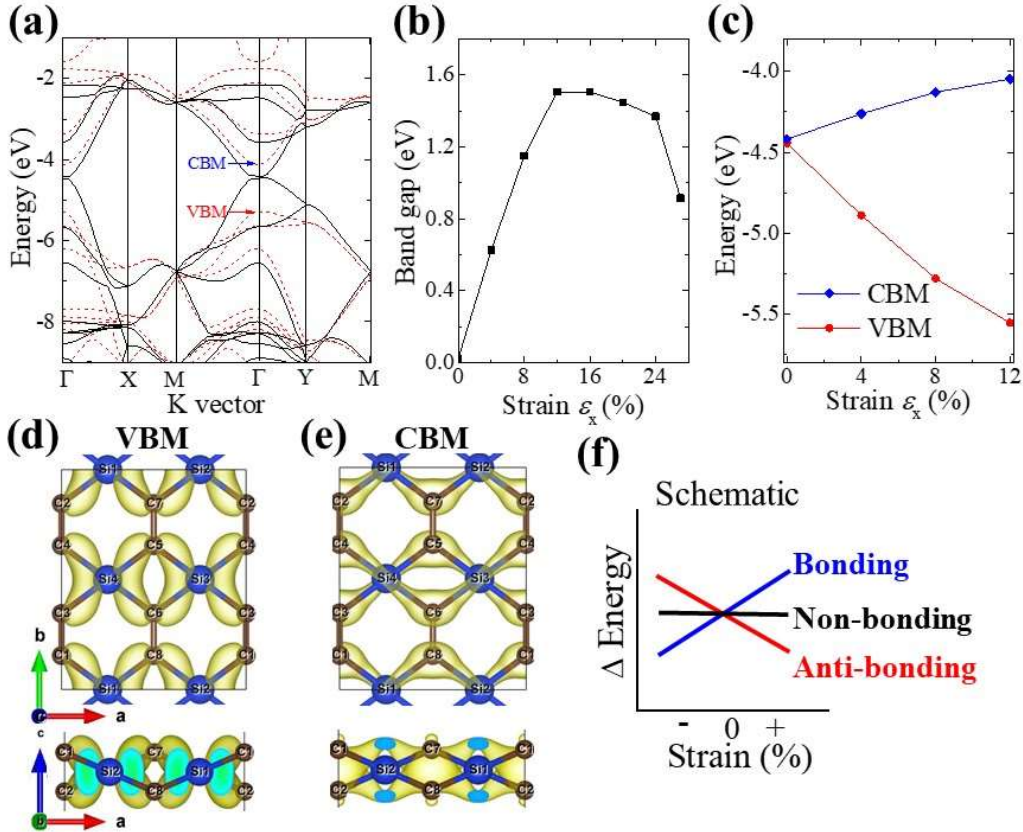

*Figure 8. Band gap opening through strain application in TH-SiC$_2$. (a) The HSE band structure without strain (solid lines) and with 8% uniaxial strain applied in the zigzag direction (dashed lines), (b) band gap as a function of uniaxial strain, (c) energy variation of the VBM and CBM with strain, respectively. All band gaps in (b) are direct gap. Energies are reference to vacuum level. Electron density contour plots of (d) VBM and (e) CBM at the isosurface value of 0.004 e/Bohr$^3$. (f) A schematic of energy response to axial strain for three typical cases of bonding, non-bonding, and anti-bonding.*

Similarly, the band structure and gap in TH-SiC is also effectively manipulated via strain as shown in Fig. 9. Since this structure is not stable with uniaxial strain applied in the *x*-axis, we only report the results of biaxial and axial strain in the *y* axis. For the uniaxial strain in the *y* direction as demonstrated in Fig. 9(a), it is found that the band gap remains indirect. However, both CBM and VBM shift to different K vectors. Without strain, the VBM and CBM in SiC are located at X and Y, respectively. With the application of 12% uniaxial strain in the *y* direction (dashed lines in Fig. 9(a)), the VBM moves away from X toward a K-vector along Γ-X, and CBM shifts from Y toward Γ. A similar story occurs for the biaxial strain as shown in Fig. 9(b), where the VBM locates nearby the K-vector of M and CBM is in the middle point along Γ-Y. The band gap variation with strain is presented in Fig. 9(c). All of the calculated band gaps are indirect and there is no indirect-direct band gap transition observed. Fig. 9(c) also suggests that the biaxial strain is more effective to tune the band gap compared to the uniaxial strain.

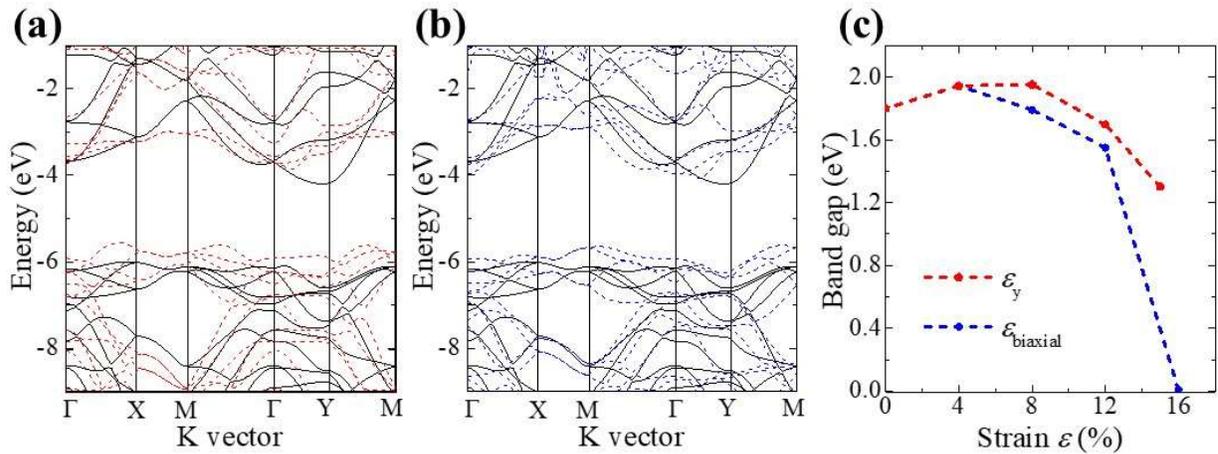

*Figure 9. Band structure and gap variation with strain in the 2D TH-SiC nanosheet. The HSE band structure without strain (solid lines) and with (a) 12% uniaxial strain in the armchair direction (dashed lines), (b) 12% biaxial strain (dashed lines). Energy is referenced to vacuum level. (c) The HSE band gap in the 2D SiC as a function of strain and all gaps are indirect within the range with phonon stability.*

### 4. Summary

Employing first-principles DFT calculations, we propose a new 2D silicon carbide nanostructure. We find the nanosheet is thermodynamically and mechanically stable. The relatively low cohesive and formation energies suggest a possibility of experimental synthesis of the materials. The new structures exhibit strong anisotropicity though the exploration of strain-stress relation and directional dependency of mechanical moduli with superior ductility. TH-SiC$_2$ structure has a tiny direct band gap with a value of 0.02 eV. However, the band gap can readily open up through different strategies including hydrogenation and strain application. The gap values can be tuned in the range of 0.02 ~ 1.72 eV, and the gap nature of direct/indirect can be manipulated. On the other hand, a closely related TH-SiC structure demonstrates an indirect band gap of 1.80 eV and strain is found to be effectively tune the gap between 0.0 ~ 1.95 eV. The ductility, anisotropicity, tunable band gap and direct/indirect gap nature in these 2D TH-SiC$_2$ and SiC may have potential applications in nanomechanics and nanoelectronics.

## Acknowledgement

This work is financially supported by the Fundamental Research Funds for the Central Universities, the Natural Science Foundation of China (Grant Nos.: 11965005), the 111 Project (B17035), and the Natural Science Basic Research plan in Shaanxi Province of China (Grant Nos.: 2020JM-186). The authors thank Arizona State University Advanced Computing Center for providing computing resources (Agave Cluster), and the computing facilities at High Performance Computing Center of Xidian University.